\documentclass[10pt,a4paper]{iopart}
\usepackage{amssymb}
\usepackage{wrapfig}
\usepackage{graphicx,caption}
\usepackage[usenames]{color}
\usepackage{cite}

\newcommand{\ket}[1]{\ensuremath{\left|#1\right\rangle}}

\begin{document}

\title[Quantum computation using quasi-hidden molecular degrees of freedom]{A robust framework for quantum computation using quasi-hidden molecular degrees of freedom}
\author{Martin Zeppenfeld}
\ead{martin.zeppenfeld@quantum-molecules.de}
\address{unaffiliated}
\date{\today, PREPRINT v0}


\begin{abstract}
We discuss a novel approach to quantum information processing with molecules based on molecular degrees of freedom which are isolated from the environment as well as from the rest of the molecule. Such a degree of freedom can provide long-term quantum storage even in a noisy environment, and provides an independent protected quantum memory while quantum operations are performed between the rest of the molecule and external systems. We present several possibilities for realizing a quasi-hidden degree of freedom in a molecule, and discuss a number of examples for using such a degree of freedom in practice. Using quasi-hidden degrees of freedom could substantially improve the prospects for a molecule-based quantum computer.
\end{abstract}



\maketitle

\section{Introduction}

Molecules cooled to very low temperatures~\cite{Prehn16,Anderegg18,Caldwell19,Ding20,Langin21,Schindewolf22} have been suggested for a wide variety of applications in quantum science, ranging from precision measurements~\cite{ACME18,Prehn21,Grasdijk21} to cold collision studies~\cite{Segev19,Nichols22,Koller22,Tang23}. A particular appealing possibility would be to use molecules as a basis for quantum information processing (QIP)~\cite{DeMille02,Andre06,Kuznetsova11,Zeppenfeld17,Wang22,Tesch02,Holland22,Bao22}. Advantageous features for this purpose include the large number and variety of internal degrees of freedom, the ability to control these internal degrees of freedom at a single quantum level, and the presence among these degrees of freedom of extremely long-lived molecular states. Moreover, for polar molecules, additional benefits include strong interactions with static electric and microwave electromagnetic fields, and long-range interactions with external quantum systems based on electric dipole-dipole interactions.

Based on the many advantageous features of quantum controlled molecules, a variety of architectures have been suggested to harness molecules for QIP. A string of polar molecules could be trapped in an optical standing wave, with dipole-dipole interactions allowing quantum gates to be performed between neighboring molecules~\cite{DeMille02}. The strong coupling of rotational transitions to microwave radiation would allow to form quantum hybrid systems combining polar molecules with superconducting qubits~\cite{Andre06}. Similarly large electric dipole moments would allow to couple polar molecules to Rydberg atoms~\cite{Kuznetsova11,Zeppenfeld17,Wang22}. Even realizing an entire quantum processor in a single large molecule has been proposed, using individual vibrational degrees of freedom as individual qubits~\cite{Tesch02}. First experiments have achieved entanglement of pairs of molecules based on dipole-dipole interactions~\cite{Holland22,Bao22}. However, given the young age of the field of cold and ultracold molecules, substantial potential remains to identify new possibilities to leverage the unique features of polar molecules for QIP.

In this paper, we consider how a molecular degree of freedom which is strongly isolated from the environment and from the remaining molecular degrees of freedom can be used to considerably enhance QIP with molecules. We begin by sketching the properties of a molecule with an idealized quasi-hidden degree of freedom and how this can be advantageous in section~\ref{sec:overview}. This is followed by a formal definition for an ideal quasi-hidden degree of freedom based on properties of the molecular Hilbert space and the molecule Hamiltonian, as well as the state operations the molecule Hamiltonian is required to allow in section~\ref{sec:formaldef}. In section~\ref{sec:realization}, we provide several examples for how an approximate version of the idealized quasi-hidden degree of freedom can be realized in a real molecule.

In section~\ref{sec:toolbox}, we consider how various quantum operations might be performed based on a specific realization of a quasi-hidden degree of freedom. On the one hand, operations are required which allow to interact with the quasi-hidden degree of freedom. On the other hand, we show that a full set of quantum operations on the non-hidden molecular degrees of freedom is possible which leave the state of the quasi-hidden degree of freedom unaffected. This includes state initialization and readout as well as entanglement with an external quantum system. Finally, in section~\ref{sec:gateexample}, we describe a key element for realizing a quantum computer based on molecules with a quasi-hidden degree of freedom. Specifically, based on Ref.~\cite{Gottesman99}, we show that entanglement between the non-hidden degrees of freedom in two molecules is a sufficient quantum resource to perform a quantum gate between two qubits stored in the quasi-hidden degrees of freedom, with only local quantum operations on each of the two molecules. We elucidate the realization of the quantum protocol step by step for the specific realization of a quasi-hidden degree of freedom, with specific molecular states.

\section{Overview}\label{sec:overview}
\begin{wrapfigure}{r}{0.45\textwidth}
	\flushright
	\captionsetup{margin={10pt,0pt}}
	\includegraphics[width=0.44\textwidth]{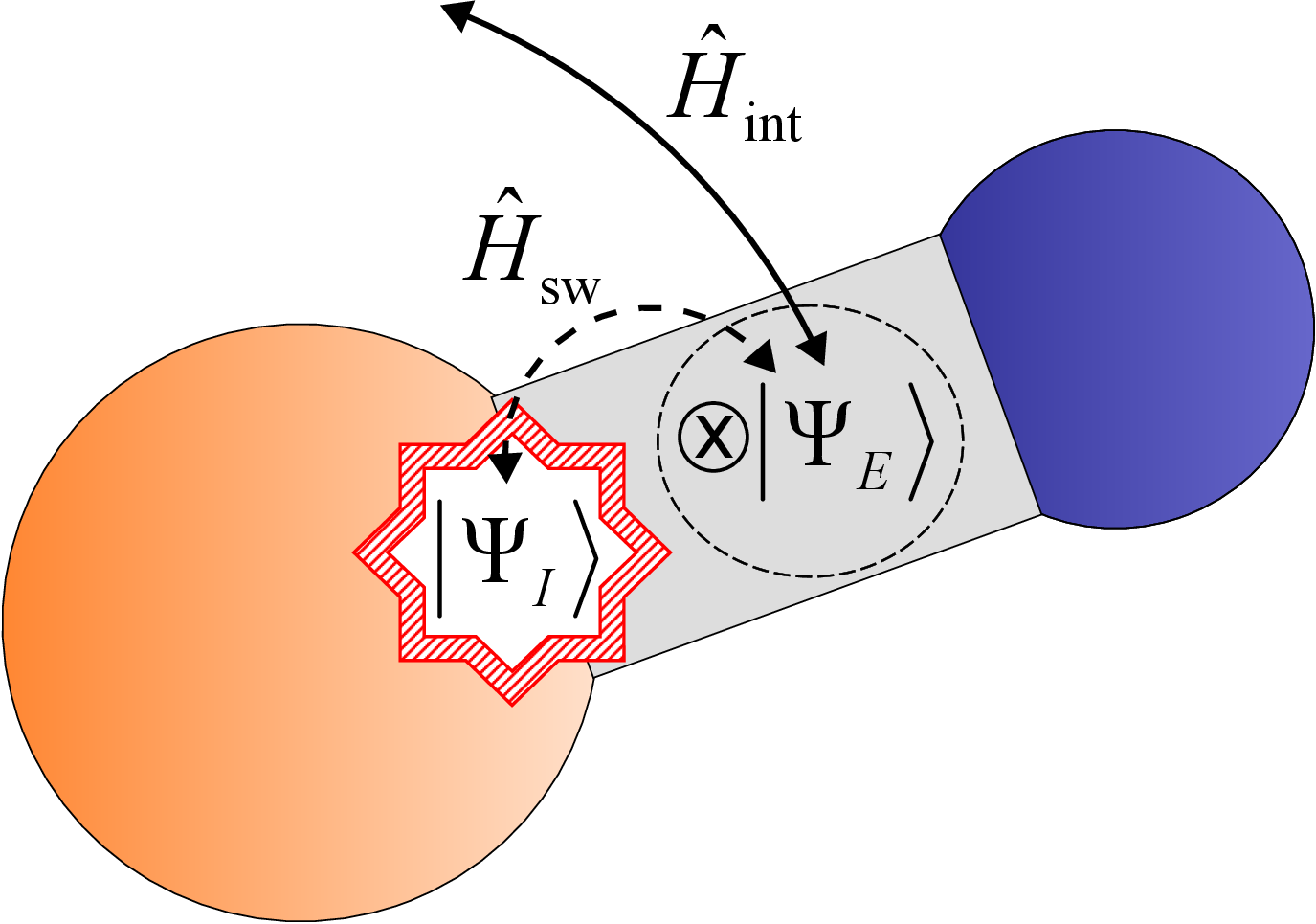}
	\caption{Illustration of a molecule with a quasi-hidden degree of freedom.}\label{fig:hiddenfreedom}
\end{wrapfigure}
In an idealized version of a molecule with a quasi-hidden degree of freedom, illustrated in Fig.~\ref{fig:hiddenfreedom} (the figure notation is explained in section~\ref{sec:formaldef}), a molecule has a degree of freedom which does not interact with the environment, and interactions with the other molecular degrees can be switched on and off at will. As a hypothetical example of such a molecule, we might consider a molecule where the electric and magnetic moments of one of the nuclei can be set to arbitrary values at will (in particular, also to zero), and where the nucleus only interacts with the intramolecular electromagnetic fields, but not with external electromagnetic fields. In many ways, such a nucleus would be an ideal quantum storage. With the electric and magnetic properties of the nucleus set to zero, the nuclear spin would exist independent of the rest of the universe, and the quantum state associated with the spin would persist unaltered indefinitely. In contrast, for nonzero nuclear moments, the nucleus would interact with the rest of the molecule. This would allow initialization and readout of the nuclear state, and moreover, to perform quantum operations between the nucleus and the rest of the molecule. Mediated by the rest of the molecule, the nucleus could in principle be entangled with an external system.

However, a key remaining challenge if a quasi-hidden degree of freedom were realized in a molecule as described, is the great difficulty in realizing high fidelity quantum interactions with systems external to a given molecule, for example, realizing a quantum gate with an external qubit. However, having a quantum degree of freedom which can be isolated from the rest of the molecule provides major advantages for solving this challenge. Thus, when the interactions with the quasi-hidden degree of freedom are switched off, the state of the rest of the molecule could be manipulated at will, without worrying about affecting the quasi-hidden degree of freedom. This would include state initialization, read out, and entanglement operations with external quantum systems, as well as physical transport of the molecule.

The ability to isolate the quasi-hidden degree of freedom would allow circumventing major imperfections when attempting to entangle the rest of the molecule with an external system. First, suppose high fidelity heralded entanglement with an external system is possible, but with a low success probability. State initialization, entanglement, and readout of the herald could be carried out repeatedly until the herald is positive, without affecting quantum information stored in the quasi-hidden degree of freedom. Second, suppose only low fidelity entanglement with an external system is possible. The entanglement operation might be carried out repeatedly, with the result stored in different non-hidden molecular degrees of freedom. Subsequently entanglement distillation might be used to realize a single high fidelity entangled state with the external system. Again, the quasi-hidden degree of freedom would remain unaffected during the process.

Once high fidelity entanglement between the rest of the molecule and an external system is achieved, the interactions with the quasi-hidden degree of freedom could be switched on. The entanglement with the external system can then be used as a quantum resource to perform quantum operations on the quantum state stored in the quasi-hidden degree of freedom, based entirely on local intramolecular interactions.

\section{Formal definition}\label{sec:formaldef}
The previous discussion provides on overview of the properties and potential benefits of a molecule with a quasi-hidden degree of freedom. Based on this, we provide a more formal definition for such a system as follows. First, it must be possible to factorize the molecular Hilbert space or (more typically) a subspace of the molecular Hilbert space, denoted by $M$
, such that it can be written as the tensor product of two smaller Hilbert spaces $I$ and $E$, i.e.,
\begin{equation}
M=I\otimes E.
\end{equation}
$I$ is meant to describe the quasi-hidden degree of freedom, and $E$ describes the remaining non-hidden molecular degrees of freedom. In the case of the molecule with a ``magic nucleus" as in section~\ref{sec:overview}, $I$ would be the Hilbert space of the nuclear spin, and $E$ would be the Hilbert space of the remaining molecule without the nuclear spin.

The conditions on the interactions between $I$, $E$, and the environment are encoded as conditions on the Hamiltonian describing the molecule and its interactions with the environment. Specifically, we require a molecule Hamiltonian $\hat{H}$ which can be written as a sum of three terms,
\begin{equation}
\hat{H}=\hat{H}_0+\hat{H}_{\rm int}+\hat{H}_{\rm sw}.
\end{equation}
Here, $\hat{H}_0$ is the molecular Hamiltonian without interactions with the environment, $\hat{H}_{\rm int}$ represents interactions with the environment, and $\hat{H}_{\rm sw}$ represents interactions between the quasi-hidden and non-hidden molecular degrees of freedom which can be switched on and off. Based on our previous discussion, $\hat{H}_0$ and $\hat{H}_{\rm int}$ only act on $E$, that is, they leave the state in $I$ unaffected. In contrast, $\hat{H}_{\rm sw}$ acts on both $E$ and $I$ in a non-separable fashion.

The previous conditions on the molecule Hamilton formulate the strong isolation of the quasi-hidden degree of freedom. However, for our system to be useful for QIP, we require the components of $\hat{H}$ to satisfy several additional conditions. First, the internal interaction Hamiltonian $\hat{H}_{\rm sw}$ must allow a full set of quantum operations between $I$ and $E$. Second, the external interaction Hamiltonian $\hat{H}_{\rm int}$ must allow a full set of quantum operations between $E$ and the environment. In particular, this must include the ability for initialization and readout of the state in $E$.

\section{Realization in a real molecule}\label{sec:realization}
So far, our discussion of a quasi-hidden molecular degree of freedom has been quite abstract, and it may be unclear how to realize such a degree of freedom in practice. In particular, our formal definition for such a degree of freedom violates the known laws of physics, since noninteracting systems are unknown to exist, and interactions can at best be approximately switched on and off at will. However, various approximate realizations of a quasi-hidden degree of freedom can be realized which convey the associated benefits for QIP.

We consider three possibilities: using opposite rotation state pairs, using chiral molecules, and using degenerate vibrational degrees of freedom, each described in detail below. The first possibility only weakly approximates an ideal quasi-hidden degree of freedom. However, this possibility is quite easy to implement in practice, and in fact, together with additional coauthors we have experimentally demonstrated the ability to observe long-lived coherences in this quasi-hidden degree of freedom in a concurrent paper~\cite{Loew23}. Moreover, this possibility exhibits most of the desirable features of a system with a quasi-hidden degree of freedom, and will be used as a point of reference for the rest of the paper.

The second possibility is an extremely good realization of a quasi-hidden degree of freedom. In fact, for certain examples of such molecules in free space, the isolation of the quasi-hidden degree of freedom is so good, that the major challenge is to engineer the internal interaction Hamiltonian $\hat{H}_{\rm sw}$. We sketch some main features, but leave the details for a future publication.

The third possibility is quite exploratory, and at this stage the details are fairly unclear to me. In many cases it is a subset of the first option, but for highly symmetric molecules it may offer substantial additional value. It is mentioned as an opportunity for future work.

\subsection{Opposite rotation state pairs under time reversal symmetry}\label{sec:rotation_pairs}
For many molecular states $\ket{\Psi}$, a distinct molecular state is obtained by applying the time reversal operator $\hat{T}$. This is easily visualized: for a molecule rotating in a given direction about some axis, applying the time reversal operator results in a molecule rotating in the opposite direction about this axis, which is a distinct state. However, many of the properties of both states are the same. The idea, then, is to consider the degree of freedom associated with replacing $\ket{\Psi}$ by $\hat{T}\ket{\Psi}$ as a quasi-hidden degree of freedom. Since $\hat{T}$ commutes with the molecule Hamiltonian, this degree of freedom is invisible to the molecule Hamiltonian. The same is true for any molecule interaction which commutes with $\hat{T}$. While molecule interactions clearly exist which don't commute with $\hat{T}$, magnetic-field interactions for example, focusing on interactions which do approximates a quasi-hidden degree of freedom.

One aspect the discussion so far has overlooked, is that $\ket{\Psi}$ and $\hat{T}\ket{\Psi}$ aren't necessarily distinct molecular eigenstates. However, it is easily shown that many such pairs of distinct degenerate energy eigenstates exist by including a component of the total molecular angular moment $J_z$ into the consideration. $\hat{J}_z$ also commutes with the molecule Hamiltonian without external interactions just like $\hat{T}$, but at the same time anticommutes with $\hat{T}$. As a result, for any simultaneous eigenstate $\ket{\Psi}$ of $\hat{J}_z$ and $\hat{H}_0$ with nonzero eigenvalue of $\hat{J}_z$, $\ket{\Psi}$ and $\hat{T}\ket{\Psi}$ are distinct but degenerate molecular states. Moreover, these states are indistinguishable for any molecule interactions which also commute with both $\hat{J}_z$ and $\hat{T}$. To approximate a quasi-hidden degree of freedom, we thus need to focus on molecule interactions which also commute with $\hat{J}_z$. In section~\ref{sec:toolbox}, we discuss in detail how the state pairs $\ket{\Psi}$ and $\hat{T}\ket{\Psi}$ can approximate the requirements for a quasi-hidden degree of freedom from section~\ref{sec:formaldef}.

\subsection{Chiral molecules}
A nearly ideal version of a quasi-hidden molecular degree of freedom can be realized by encoding the hidden degree of freedom in the opposite enantiomers of a chiral molecule. For a chiral molecule in free space, the two enantiomers are almost indistinguishable based on external field interactions. Both enantiomers have an identical rovibronic energy level structure and identical matrix elements for a wide range of external field interactions. In fact, distinguishing enantiomers is such a difficult task that considerable effort has been directed at developing techniques to do so~\cite{He11,Patterson13a,Fanood15}. For a chiral molecule with a permanent electric dipole moment directed along all three principle molecular axes of inertia, the enantiomers can be distinguished, for example by three wave mixing~\cite{Patterson13b}. However, for molecules with a permanent dipole moment directed along at most two of the principle axes of inertia, the two enantiomers are indistinguishable based on electromagnetic dipole interactions. As a result, if a quantum superposition between the two enantiomers were to exist, this would persist without decoherence independent of any manipulation of the molecule based on electric or magnetic dipole interactions. This would even include dipolar spontaneous radiative decay of excited molecular states. The molecular state (excluding the hidden degree of freedom encoded in the molecular handedness) could thus be read out and initialized based on optical pumping (provided suitable optical pumping transitions exist), while a quantum superposition encoded in the handedness of the molecule would continue a blissful coherent existence, oblivious to anything going on in the rest of the molecule.

The handedness of a chiral molecule is such a good hidden degree of freedom, that the major challenge using this as a resource for QIP would be engineering the internal interaction Hamiltonian $\hat{H}_{\rm sw}$. An interaction Hamiltonian allowing universal manipulation of the hidden quantum state would require two components, the ability to interconvert the two enantiomers, and the ability to distinguish the two enantiomers. Describing the hidden quantum state as points on a Bloch sphere, with spin up/down along the $z$-axis corresponding to a pure left- or right-handed enantiomer, distinguishing the enantiomers would correspond to the $\hat{s}_z$ operator, and interconverting the enantiomers would correspond to, e.g., the $\hat{s}_x$ operator.

Interconverting enantiomers in a quantum coherent fashion follows a fairly obvious blueprint. A molecule would need to be chosen such that tunneling between the enantiomers happens in highly excited vibrational states~\cite{Quack06}, without dissociation of the molecule (electronically excited states could also be considered~\cite{Quack86}). Exciting the molecule to high vibrational states would allow the interconversion component of the internal interaction Hamiltonian $\hat{H}_{\rm sw}$ to be ``switched on" essentially in the ideal sense of section~\ref{sec:formaldef}. Care would need to be taken, however, to avoid losing track of the molecular state in the sea of other molecular vibrational states during the excitation process.

Distinguishing enantiomers based on external field interactions is a substantially more difficult task. For a permanent electric dipole moment along at most two of the principle axes of inertia, this is only possible via higher order multipole interactions. The details will be explored in a future publication.

\subsection{Degenerate vibrational degrees of freedom}
A final possibility to realize a quasi-hidden molecular degree of freedom is to use degenerate vibrational states. For molecules with a 3-fold or higher rotation axis, many vibrational states come in degenerate pair, and these pairs can be considered as a quasi-hidden degree of freedom. However, in the case of a single 3-fold or higher rotation axis, the degenerate vibrational state pairs are simply mirror images under time reversal symmetry, and thus fall under the discussion in section~\ref{sec:rotation_pairs}.

For molecules with tetrahedral, octahedral and icosahedral symmetry, vibrational states exist which are triply degenerate, and in the case of icosahedral symmetry, even 4-fold or 5-fold degenerate. Clearly this degeneracy is not generated by time reversal symmetry, since time reversal symmetry can only account for a 2-fold degeneracy. It thus seems likely that degenerate vibrational states in molecules with tetrahedral or higher symmetry could be used as a separate category of a quasi-hidden degree of freedom. In this context, two important questions would need to be answered. First, to what degree are the vibrational states truly degenerate when taking into account coupling to the molecular rotation, and to what degree are superpositions of these states impervious to common interactions of the molecule with the environment? For example, is the degeneracy lifted when the molecule is subjected to an electric or magnetic field? Second, how can the internal interaction Hamiltonian $\hat{H}_{\rm sw}$ be engineered, allowing interactions with the quasi-hidden degree of freedom?

\section{Demonstration that opposite rotation state pairs fulfill the requirements for a quasi-hidden degree of freedom}\label{sec:toolbox}
In this section, we demonstrate how using opposite rotation state pairs under time reversal symmetry at least approximately satisfies all the requirements of a quasi-hidden degree of freedom defined in section~\ref{sec:formaldef}. In particular, we identify a possible choice of Hilbert subspaces $I$, $E$, and $M$, we discuss how this choice (approximately) satisfies the requirements on $\hat{H}_0$, $\hat{H}_{\rm int}$, and $\hat{H}_{\rm sw}$ in terms of their action on $I$ and $E$, and we sketch how the various quantum operations which were identified might be realized.

For our discussion, we restrict ourselves to the rotational structure of a simple dipolar $\Sigma_0$ molecule, described by rotational quantum numbers $J$ and $m$. $J$ is the total angular momentum, and $m$ is the projection of $J$ on a space-fixed axis. 
We define $M$ as the Hilbert space generated by the set of states with $m\ne0$. We define $E$ as the quotient space of $M$ under the equivalence relation which identifies every state $\ket{J,m}$ with the state $\hat{T}\ket{J,m}=\ket{J,-m}$, i.e. the Hilbert space generated by the states $\ket{J,m}$ in $M$ where the states $\ket{J,m}$ and $\ket{J,-m}$ are considered identical. Finally, we define $I$ as the two-dimensional Hilbert space describing the sign of $m$. $M$ is then clearly isomorphic with $I\otimes E$.

Since the energy of a molecular state in free space does not depend on $m$, the molecule Hamiltonian $\hat{H}_0$ clearly acts as the identity operator on $I$. Concerning $\hat{H}_{\rm int}$ and $\hat{H}_{\rm sw}$, we can can define $\hat{H}_{\rm int}$ as the sum of all interactions with the environment which do not affect the state in $I$, and $\hat{H}_{\rm sw}$ as all remaining interactions with the environment. While this definition satisfies the requirements that $\hat{H}_{\rm int}$ does not affect the state in $I$, it doesn't completely satisfy the requirement that $\hat{H}_{\rm sw}$ can be switched on and off. Certain interactions with the environment, for example spontaneous radiative decay, will affect the state in $I$ but can never be switched off entirely. However, in practice these interactions can be very weak, and they would need to be regarded as a residual source of decoherence.

The final requirement is that $\hat{H}_{\rm int}$ and $\hat{H}_{\rm sw}$ can affect a full set of quantum operations on the states in $E$ and $I$. For this purpose, it is sufficient to consider the effect of the molecule dipole operator $\hat{\mathbf{d}}$, consisting of the three projections $\hat{d}_x$, $\hat{d}_y$, and $\hat{d}_z$ of the molecule's dipole moment on space fixed axes. Defining the $z$-direction as our quantization axis, $\hat{d}_z$ leaves $m$ invariant and acts identically on positive and negative $m$. It thus acts as the identity operator on $I$, and $\hat{H}_{\rm int}$ thus includes interactions with the environment involving $\hat{d}_z$.

For usual phase conventions, $\hat{d}_x$ also acts identically on positive and negative $m$. $\hat{d}_x$ thus also acts as the identity operator on $I$, with one exception. We observe that the operator $\hat{d}_x$ has two invariant subspaces in the set of rotational states, the set of even linear combinations of positive and negative $m$, $\ket{J,m}+\ket{J,-m}$, and the set of odd linear combinations$\ket{J,m}-\ket{J,-m}$. The action of $\hat{d}_x$ on the even and odd linear combinations is identical, except for the fact that the even linear combinations include the states $m=0$, whereas the odd linear combinations do not. $\hat{H}_{\rm int}$ thus includes interactions with the environment involving $\hat{d}_x$ which do not result in coupling to any of the $m=0$ states. This can be achieved, for example, by applying an electric field to obtain a Stark splitting between different $|m|$ for a given $J$, and to resonantly couple different $|m|>0$ via $\hat{d}_x$ such that the coupling to $m=0$ is always far off resonance.

The matrix elements of $\hat{d}_y$ are almost the same as the matrix elements of $\hat{d}_x$, except that $\hat{d}_y$ couples even linear combinations of $\pm m$ to odd linear combinations. $\hat{d}_y$ thus strongly affects the state in $I$, and interactions with the environment involving $\hat{d}_y$ can be considered part of $\hat{H}_{\rm sw}$.

Based on the previous discussion, the operators $\hat{H}_0$, $\hat{d}_z$, and $\hat{d}_x$ allow to manipulate the molecular state in $E$ while leaving the state in $I$ unaffected (with proper care when using $\hat{d}_x$), whereas $\hat{d}_y$ additionally allows to manipulate the molecular state in $I$. The remaining question is if these operators allow constructing an arbitrary unitary transformation on the molecular rotational Hilbert space, i.e. to construct an arbitrary quantum operator. Proving that this is the case for arbitrary $J$ is not self evident for me. However, for neighboring $J$ states $J$ and $J+1$ with low values of $J$, it is reasonably straightforward to demonstrate that the operators $\hat{H}_0$, $\hat{d}_z$, and $\hat{d}_x$ generate all possible unitary transformations acting on the $m$-sublevels of $J$ and $J+1$ which preserve the parity of the linear combinations of $\pm m$, as I have verified numerically for $J$ up to $6$. Adding the $\hat{d}_y$ operator additionally allows to break the invariance of the even and odd linear combinations.

To demonstrate that a set of operators generates an entire group, we start with the set of operators (in our case $\hat{H}_0$, $\hat{d}_z$, and $\hat{d}_x$), take the commutator of any pair of operators in the set, and add the result to the set if it is linearly independent of the previous members. If this results in a set of linearly independent operators whose number is equal to the maximum number of linearly independent operators for the given group, this proves that the initial set of operators generates the entire group.

\subsection{Realizing quantum operations on the molecule Hilbert space in practice}
The previous discussion demonstrates that at least in theory, the molecule dipole operators $\hat{\mathbf{d}}$ combined with the noninteracting molecule Hamiltonian $\hat{H}_0$ allow arbitrary manipulations of the molecular state at least up to $J=7$. We consider how such state manipulations might be achieved in practice.

Manipulating the molecular state with external electric fields while regarding the molecule as a closed quantum system is entirely straightforward based on established techniques. Static or radiofrequency electric fields are applied along the $x$-, $y$-, or $z$-direction to couple to the $\hat{d}_x$, $\hat{d}_y$, or $\hat{d}_z$ operator. As mentioned previously, a static electric field can be used to separate different $|m|$-sublevels, allowing transitions between specific $m$-sublevels to be driven selectively by radiofrequency radiation based on a resonance condition. Alternatively, optimal control theory~\cite{Koch16} could be used to perform quantum gates on a shorter timescale.

The bigger challenge lies in coupling the molecular state in $E$ to an external quantum system, and to read out and initialize this state, without affecting the state in $I$. We envision two possibilities. A particularly appealing approach would be to couple the molecule to a Rydberg atom via dipole-dipole interactions~\cite{Kuznetsova11,Zeppenfeld17,Wang22}. The molecule and the Rydberg atom need to be brought into close proximity, and an electric field would be applied to bring a specific Rydberg atom transition into resonance with a specific molecule transition. Coupling the molecule to a specific Rydberg atom transition with a well defined orientation between the molecule and the Rydberg atom would allow controlling the polarization of the electric field experienced by the molecule due to the Rydberg atom. This would allow, e.g., to only couple to $\Delta m=0$ molecule transitions.

Coupling to a Rydberg atom allows a full set of quantum operations, as illustrated by the following specific example. We consider a Rydberg atom with states $\ket{r_1}$ and $\ket{r_2}$ which are resonant to the molecule transition from $\ket{J=2,|m|=1}$ to $\ket{J=3,|m|=1}$ for a certain applied electric field. We describe the state of the combined system as $\ket{J,|m|,r_i}$. The states $\ket{2,1,r_2}$ and $\ket{3,1,r_1}$ have the same energy, and for a close separation between the molecule and the Rydberg atom, Rabi oscillations between these states will take place based on dipole-dipole interactions.

Suppose the molecule is initially in the state $a\ket{2,1}+b\ket{3,1}+c\ket{4,3}$, with probability amplitudes $a$, $b$, and $c$. The Rydberg atom is initialized to the state $\ket{r_1}$. The initial state of the combined system is
\begin{equation}
\ket{\Psi_{\rm init}}=a\ket{2,1,r_1}+b\ket{3,1,r_1}+c\ket{4,3,r_1}.
\end{equation}
After a half period of Rabi oscillations between the states $\ket{2,1,r_2}$ and $\ket{3,1,r_1}$, the state of the combined system will be
\begin{equation}
\ket{\Psi_{\rm final}}=a\ket{2,1,r_1}+b\ket{2,1,r_2}+c\ket{4,3,r_1}.
\end{equation}
This result can be useful in three ways. First and second, if the state of the Rydberg atom is subsequently read out, and determined to be $\ket{r_2}$, the molecule will have been initialized to the state $\ket{J=2,|m|=1}$, and the previous state of the molecule will be determined to have been $\ket{J=3,|m|=1}$. Third, for $b=c=1/\sqrt{2}$, the final state of the combined system is a Bell state: the molecule and the Rydberg atom will have been entangled. We emphasize that for all these operations, the sign of $m$ is a hidden degree of freedom.

A second technologically substantially more daunting possibility for quantum coupling to a molecule would be to replace the Rydberg atom with a microwave resonator which is fully controlled at a single photon level and which is strongly coupled to the molecule~\cite{Andre06}. The microwave resonator would again be resonant to a specific molecule transition, and controlling the polarization of the microwave field again allows to control the type of coupling to the molecule, in particular to couple to $E$ while leaving the state in $I$ unaffected. Using the microwave resonator allows for readout and initialization of as well as entanglement with the molecular state, roughly analogous to the example with the Rydberg atom.

\section{Realizing a quantum gate between two hidden qubits based on entanglement between the non-hidden degrees of freedom}\label{sec:gateexample}
In this section, we provide a detailed illustration for how a quasi-hidden molecular degree of freedom might be used for a quantum computing architecture. In the previous, we have discussed two useful features of a molecule with a quasi-hidden degree of freedom. First, due to the isolation from the environment, the quasi-hidden degree of freedom allows to store quantum information with a long coherence time. Second, the isolation of the quasi-hidden degree of freedom from the rest of the molecule allows performing quantum operations between the rest of the molecule and the environment without affecting the quasi-hidden degree of freedom. We now show how these features can be combined, allowing to perform a quantum gate between two qubits stored in the quasi-hidden degree of freedom in two separate molecules, with only local single-molecule quantum interactions involving the quasi-hidden degree of freedom in either of the two molecules.

\begin{wrapfigure}{r}{0.45\textwidth}
	\flushright
	\captionsetup{margin={10pt,0pt}}
	\includegraphics[width=0.44\textwidth]{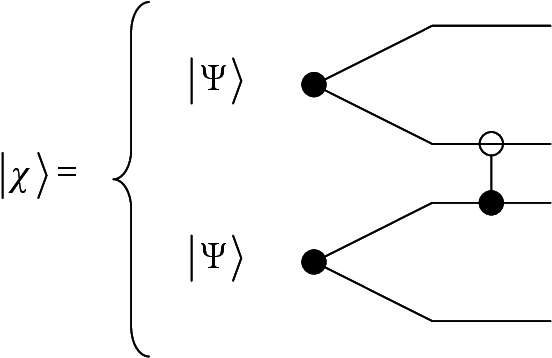}
	\caption{Preparation of the state $\ket{\chi}$ involved in the quantum protocol, involving a total of four qubits as described in the main text. $\ket{\Psi}$ refers to a Bell state between two qubits.}\label{fig:chi}
\end{wrapfigure}

We define our problem more precisely as follows. Suppose a qubit is stored in the quasi-hidden degree of freedom in two molecules. Suppose, furthermore, that a Bell state entanglement between the non-hidden degrees of freedom of the two molecules has been established without affecting the quasi-hidden degree of freedom, using the methods outlined previously in this paper. Is it possible to perform a quantum gate on the two stored qubits using the Bell state entanglement as a quantum resource, with only local quantum operations in each of the two molecules?

Our solution to the stated problem is based on a quantum protocol described in a 1999 paper by Gottesman and Chuang~\cite{Gottesman99}. We first describe this protocol and explain how this protocol results in a quantum gate between two qubits stored in the quasi-hidden degrees of freedom in two molecules in section~\ref{sec:protocol}. We then lay out a detailed realization of this protocol in a molecule with opposite-rotation state pairs as the quasi-hidden degree of freedom in section~\ref{sec:quantumgate}. In particular, this includes demonstrating that a Bell state entanglement between the two molecules is a sufficient quantum resource for realizing the quantum protocol, in contrast with the more complicated initial entangled state in the 1999 paper.

\subsection{A quantum protocol for remote realization of a quantum gate}\label{sec:protocol}
The basis for the quantum protocol is an entangled state $\ket{\chi}$ encoded in a total of four qubits. Two of these qubits are stored in the non-local degrees of freedom of each of the two molecules, with
\begin{equation}
\ket{\chi}=\frac{(\ket{00}_l+\ket{11}_l)\ket{00}_r+(\ket{00}_l-\ket{11}_l)\ket{11}_r}{2}.
\end{equation}
To clarify the notation, $\ket{00}_l$, for example, denotes the two qubits in the ``left" molecule being in the state $0$, and $\ket{11}_r$ denotes the two qubits in the ``right" molecule being in the state $1$. A possible preparation of the state $\ket{\chi}$ is illustrated in Fig.~\ref{fig:chi}. Starting with a Bell state $\ket{00}+\ket{11}$ in both of the two molecules, performing a controlled $z$ gate on the combination of one of the qubits in both molecules results in the state $\ket{\chi}$.

\begin{figure}[t]
	\centering
	\includegraphics[width=1\textwidth]{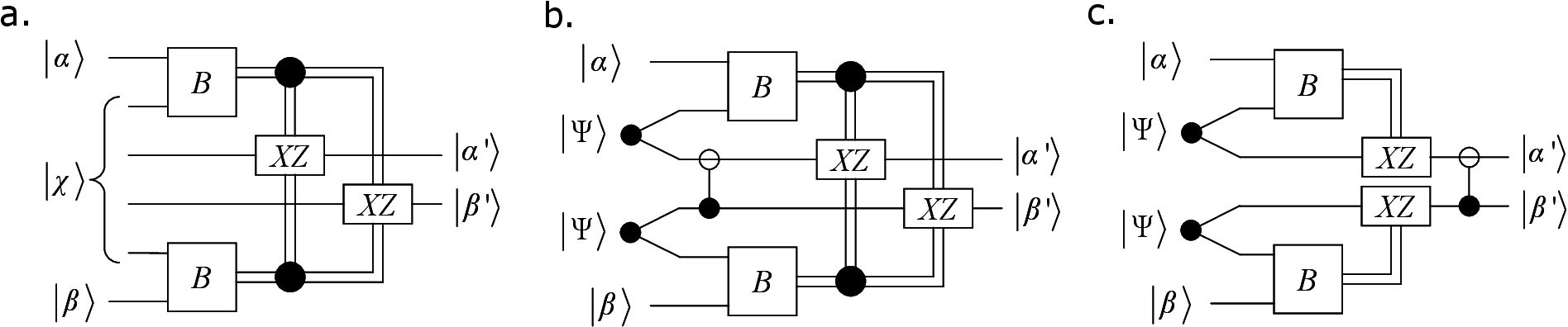}
	\caption{Quantum protocol for teleporting a quantum gate, as described in the main text. Single lines denote quantum channels, double lines denote classical channels.}\label{fig:protocol}
\end{figure}

The overall quantum protocol is illustrated in Fig.~\ref{fig:protocol}a. $\ket{\alpha}$ and $\ket{\beta}$ represent the qubits stored in the quasi-hidden degree of freedom in the two molecules. A Bell state measurement involving the quasi-hidden degree of freedom and one of the two qubits involved in the state $\ket{\chi}$ is performed in both molecules. This is followed by suitable single qubit rotations on the second qubit in the state $\ket{\chi}$ in each molecule depending on the results of the Bell state measurements. Importantly, these single qubit rotations exclusively involve swapping the states $\ket{0}$ and $\ket{1}$ and adding a $\pi$ phase shift to the state $\ket{1}$ relative to the state $\ket{0}$. Finally, the result is stored back in the quasi-hidden degree of freedom.

The effect of our quantum protocol can be understood quite easily based on Figs.~\ref{fig:protocol}b and c. In Fig.~\ref{fig:protocol}b, we simply add the preparation of the state $\ket{\chi}$ from Fig.~\ref{fig:chi} to the diagram. In Fig.~\ref{fig:protocol}c, we make use of the fact that a controlled $z$ gate followed by the previously specified single qubit rotations is equivalent to a different combination of the specified single qubit rotations followed by the same controlled $z$ gate. For suitable single qubit rotations, Fig.~\ref{fig:protocol}c is thus equivalent to Fig.~\ref{fig:protocol}a. We now chose the single qubit rotations in Fig.~\ref{fig:protocol}a such that the combination of Bell state measurement and single qubit rotations corresponds to quantum teleportation of the states $\ket{\alpha}$ and $\ket{\beta}$ onto the remaining quantum channels. Fig.~\ref{fig:protocol}c then consists of quantum teleportation of the states $\ket{\alpha}$ and $\ket{\beta}$ followed by a controlled $z$ gate on the states $\ket{\alpha}$ and $\ket{\beta}$. Fig.~\ref{fig:protocol}a is equivalent to Fig.~\ref{fig:protocol}c, except that the order of the quantum teleportation and the controlled $z$ gate is reversed: a controlled $z$ gate is effectively contained in the definition of the state $\ket{\chi}$, and via the protocol in Fig.~\ref{fig:protocol}a, this controlled $z$ gate is teleported onto the states $\ket{\alpha}$ and $\ket{\beta}$.

\subsection{Explicit realization of a remote quantum gate using opposite-rotation state pairs}\label{sec:quantumgate}
To further illustrate how a quasi-hidden degree of freedom can be used in practice, we explicitly demonstrate realization of the quantum protocol in the preceding section using opposite-rotation state pairs in two $\Sigma_0$ polar molecules as described in section~\ref{sec:toolbox}. We use basis states $\ket{J,|m|,\pm}_l\ket{J',|m'|,\pm'}_r$ to describe the internal state of the two molecules, where the $\pm$ denotes the sign of $m$, and the subscripts $l$ and $r$ again label the ``left" and the ``right" molecule. We also use the notation $\ket{\pm}_i$ and $\ket{J,|m|}_i$ to denote the quasi hidden degree of freedom and the non-hidden degrees of freedom of a given molecule separately.

We assume the molecules to initially be in the state
\begin{eqnarray}\label{initialstate}
\Psi_{\rm init}=\frac{1}{\sqrt{2}}\times&\Big(\alpha_+\ket{+}_l+\alpha_-\ket{-}_l\Big)\otimes\nonumber\\
&\Big(\ket{2,1}_l\ket{2,1}_r+\ket{3,1}_l\ket{3,1}_r\Big)\otimes\nonumber\\
&\Big(\beta_+\ket{+}_r+\beta_-\ket{-}_r\Big).
\end{eqnarray}
Here, $\alpha_\pm$ and $\beta_\pm$ are probability amplitudes for the states $\ket{\alpha}$ and $\ket{\beta}$. Thus, the states $\ket{\alpha}$ and $\ket{\beta}$ are initially stored in the quasi-hidden degree of freedom of the two molecules, and the non-hidden degrees of freedom of the two molecules are entangled as a simple Bell state.

To realize the quantum protocol from the preceding section, we require the ability to perform a number of single molecule operations on the two molecules. Following the discussion in section~\ref{sec:toolbox}, we assume that applying a static electric field results in a Stark splitting between different $|m|$-sublevels, allowing to drive each transition from a state $\ket{J,|m|}$ to $\ket{J+1,|m'|}$ with $|m|-|m'|=0,\pm1$ in isolation. Moreover, we assume that using radiofrequency radiation with $x$-polarization allows to drive a transition from $\ket{J,|m|}$ to $\ket{J+1,|m|\pm1}$ without affecting the state $\ket{\pm}$, whereas using circular polarization allows to drive the transitions $\ket{J,|m|,+}$ to $\ket{J+1,|m|\pm1,+}$ and $\ket{J,|m|,-}$ to $\ket{J+1,|m|\pm1,-}$ independently.

The first step to realize our quantum protocol is to convert the Bell state in Eq.~\ref{initialstate} into the state $\ket{\chi}$ from the preceding section. For this purpose, we need to identify the two qubits stored in the non-hidden degrees of freedom in each molecule. This is done via a state mapping between four rotational molecular states and the basis states for the two qubits. Specifically, we associate
\begin{eqnarray}\label{statemapping}
\ket{1,1}\leftrightarrow\ket{01}\nonumber\\
\ket{2,1}\leftrightarrow\ket{00}\nonumber\\
\ket{3,1}\leftrightarrow\ket{11}\nonumber\\
\ket{4,1}\leftrightarrow\ket{10},
\end{eqnarray}
where the left hand side is in the $\ket{J,|m|}$ basis and the right hand side is in the two-qubit basis. If we now perform a $\pi/2$ rotation between the state $\ket{2,1}$ and the state $\ket{3,1}$ in one of the two molecules such that the state $\ket{2,1}$ becomes $(\ket{2,1}+\ket{3,1})/\sqrt{2}$ and the state $\ket{3,1}$ becomes $(\ket{2,1}-\ket{3,1})/\sqrt{2}$, the state of our pair of molecule becomes
\begin{eqnarray}
\Psi_1=\frac{1}{2}\times\Big(\alpha_+\ket{+}_l+\alpha_-\ket{-}_l\Big)\otimes\nonumber\\
\Big(\ket{2,1}_l\ket{2,1}_r+\ket{2,1}_l\ket{3,1}_r+\ket{3,1}_l\ket{2,1}_r-\ket{3,1}_l\ket{3,1}_r\Big)\otimes\nonumber\\
\Big(\beta_+\ket{+}_r+\beta_-\ket{-}_r\Big).
\end{eqnarray}
In combination with the state mapping in Eq.~\ref{statemapping}, we see that the quantum state stored in the non-hidden molecular degrees of freedom is precisely the state $\ket{\chi}$. We have thus demonstrated that a simple Bell state is a sufficient basis for our quantum protocol.

The next step of our quantum protocol is to perform a Bell state measurement on the combination of the quasi-hidden degree of freedom and one of the two qubits stored in the non-hidden degrees of freedom. However, our requirements from section~\ref{sec:formaldef} allow us to perform a measurement on the non-hidden degrees of freedom, but not on the quasi-hidden degree of freedom. Moreover, it is unclear how to perform a measurement of one of the two non-hidden qubits without affecting the other. As an alternative, we first swap the quasi-hidden qubit with one of the two non-hidden qubits. This allows us to perform the Bell state measurement with the non-hidden degrees of freedom whereas the quasi-hidden degree of freedom needs to be left unaffected by the measurement.

Swapping the quasi-hidden qubit with one of the non-hidden qubits is performed with the following mapping:
\begin{eqnarray}\label{stateswap}
\ket{1,1,+}\leftrightarrow\ket{01;0}\rightarrow\ket{00;1}\leftrightarrow\ket{2,1,-}\nonumber\\
\ket{1,1,-}\leftrightarrow\ket{01;1}\rightarrow\ket{01;1}\leftrightarrow\ket{1,1,-}\nonumber\\
\ket{2,1,+}\leftrightarrow\ket{00;0}\rightarrow\ket{00;0}\leftrightarrow\ket{2,1,+}\nonumber\\
\ket{2,1,-}\leftrightarrow\ket{00;1}\rightarrow\ket{01;0}\leftrightarrow\ket{1,1,+}\nonumber\\
\ket{3,1,+}\leftrightarrow\ket{11;0}\rightarrow\ket{10;1}\leftrightarrow\ket{4,1,-}\nonumber\\
\ket{3,1,-}\leftrightarrow\ket{11;1}\rightarrow\ket{11;1}\leftrightarrow\ket{3,1,-}\nonumber\\
\ket{4,1,+}\leftrightarrow\ket{10;0}\rightarrow\ket{10;0}\leftrightarrow\ket{4,1,+}\nonumber\\
\ket{4,1,-}\leftrightarrow\ket{10;1}\rightarrow\ket{11;0}\leftrightarrow\ket{3,1,+}.
\end{eqnarray}
The first and last column correspond to the $\ket{J,|m|,\pm}$ basis, the central two columns correspond to a three qubit basis. We thus need to swap the population between the states $\ket{2,1,-}$ and $\ket{1,1,+}$ and between the states $\ket{3,1,+}$ and $\ket{4,1,-}$ with the population in the other states remaining unaffected. However, since the states $\ket{1,1,+}$ and $\ket{4,1,-}$ have no overlap with the state $\Psi_1$, we only need to transfer the population from the first to the second state in both cases. In the first case, this can be achieved with a $\pi$ rotation between the state $\ket{2,1,-}$ and the state $\ket{1,0}$ (the $\pm$ label doesn't exist for the states $m=0$), followed by a $\pi$ rotation between the state $\ket{1,0}$ and the state $\ket{2,0}$, followed by a $\pi$ rotation between the state $\ket{2,0}$ and the state $\ket{1,1,+}$. In the second case, this can be achieved with a $\pi$ rotation between the state $\ket{3,1,+}$ and the state $\ket{4,0}$, followed by a $\pi$ rotation between the state $\ket{4,0}$ and the state $\ket{3,0}$, followed by a $\pi$ rotation between the state $\ket{3,0}$ and the state $\ket{4,1,-}$. Following these operations, the state of the pair of molecules is given by
\begin{eqnarray}
\Psi_2=\frac{1}{2}\times\bigg(\nonumber\\
\fl\alpha_+\beta_+\Big(\ket{+}_l\ket{2,1}_l\ket{2,1}_r\ket{+}_r+\ket{+}_l\ket{2,1}_l\ket{4,1}_r\ket{-}_r+\nonumber\\
\ket{-}_l\ket{4,1}_l\ket{2,1}_r\ket{+}_r-\ket{-}_l\ket{4,1}_l\ket{4,1}_r\ket{-}_r\Big)+\nonumber\\
\fl\alpha_+\beta_-\Big(\ket{+}_l\ket{2,1}_l\ket{1,1}_r\ket{+}_r+\ket{+}_l\ket{2,1}_l\ket{3,1}_r\ket{-}_r+\nonumber\\
\ket{-}_l\ket{4,1}_l\ket{1,1}_r\ket{+}_r-\ket{-}_l\ket{4,1}_l\ket{3,1}_r\ket{-}_r\Big)+\nonumber\\
\fl\alpha_-\beta_+\Big(\ket{+}_l\ket{1,1}_l\ket{2,1}_r\ket{+}_r+\ket{+}_l\ket{1,1}_l\ket{4,1}_r\ket{-}_r+\nonumber\\
\ket{-}_l\ket{3,1}_l\ket{2,1}_r\ket{+}_r-\ket{-}_l\ket{3,1}_l\ket{4,1}_r\ket{-}_r\Big)+\nonumber\\
\fl\alpha_-\beta_-\Big(\ket{+}_l\ket{1,1}_l\ket{1,1}_r\ket{+}_r+\ket{+}_l\ket{1,1}_l\ket{3,1}_r\ket{-}_r+\nonumber\\
\ket{-}_l\ket{3,1}_l\ket{1,1}_r\ket{+}_r-\ket{-}_l\ket{3,1}_l\ket{3,1}_r\ket{-}_r\Big)\bigg).
\end{eqnarray}

We now wish to perform a measurement of the molecule state in the non-hidden degrees of freedom in the Bell basis, given by the four states $(\ket{00}\pm\ket{11})/\sqrt{2}$ and $(\ket{01}\pm\ket{10})/\sqrt{2}$ in the two-qubit basis, or the states $(\ket{2,1}\pm\ket{3,1})/\sqrt{2}$ and $(\ket{1,1}\pm\ket{4,1})/\sqrt{2}$ in the $\ket{J,|m|}$ basis. As an alternative, we can first transform the Bell states into pure $\ket{J,|m|}$ states. This is achieved via $\pi/2$ rotations such that
\begin{eqnarray}\label{pihalfrotation}
\ket{2,1}\rightarrow\frac{1}{\sqrt{2}}\Big(\ket{2,1}+\ket{3,1}\Big)\nonumber\\
\ket{3,1}\rightarrow\frac{1}{\sqrt{2}}\Big(\ket{2,1}-\ket{3,1}\Big)\nonumber\\
\ket{1,1}\rightarrow\frac{1}{\sqrt{2}}\Big(\ket{1,1}+\ket{1,1}\Big)\nonumber\\
\ket{4,1}\rightarrow\frac{1}{\sqrt{2}}\Big(\ket{1,1}-\ket{4,1}\Big).
\end{eqnarray}
In the first two cases, this is achieved by directly coupling the two states via a $\pi/2$ pulse. In the last two cases, this can be achieved by using the states $\ket{2,2}$ and $\ket{3,2}$ as intermediary states.

Following the previous transformation, the state of the pair of molecule is given by
\begin{eqnarray}\label{finalstate}
\Psi_{\rm final}=\frac{1}{4}\times\bigg(\nonumber\\
\fl\ket{1,1}_l\ket{1,1}_r\Big(\alpha_-\beta_-\ket{+}_l\ket{+}_r+\alpha_-\beta_+\ket{+}_l\ket{-}_r+\alpha_+\beta_-\ket{-}_l\ket{+}_r-\alpha_+\beta_+\ket{-}_l\ket{-}_r\Big)+\nonumber\\
\fl\ket{1,1}_l\ket{2,1}_r\Big(\alpha_-\beta_+\ket{+}_l\ket{+}_r+\alpha_-\beta_-\ket{+}_l\ket{-}_r+\alpha_+\beta_+\ket{-}_l\ket{+}_r-\alpha_+\beta_-\ket{-}_l\ket{-}_r\Big)+\nonumber\\
\fl\ket{1,1}_l\ket{3,1}_r\Big(\alpha_-\beta_+\ket{+}_l\ket{+}_r-\alpha_-\beta_-\ket{+}_l\ket{-}_r+\alpha_+\beta_+\ket{-}_l\ket{+}_r+\alpha_+\beta_-\ket{-}_l\ket{-}_r\Big)+\nonumber\\
\fl\ket{1,1}_l\ket{4,1}_r\Big(\alpha_-\beta_-\ket{+}_l\ket{+}_r-\alpha_-\beta_+\ket{+}_l\ket{-}_r+\alpha_+\beta_-\ket{-}_l\ket{+}_r+\alpha_+\beta_+\ket{-}_l\ket{-}_r\Big)+\nonumber\\
\fl\ket{2,1}_l\ket{1,1}_r\Big(\alpha_+\beta_-\ket{+}_l\ket{+}_r+\alpha_+\beta_+\ket{+}_l\ket{-}_r+\alpha_-\beta_-\ket{-}_l\ket{+}_r-\alpha_-\beta_+\ket{-}_l\ket{-}_r\Big)+\nonumber\\
\fl\ket{2,1}_l\ket{2,1}_r\Big(\alpha_+\beta_+\ket{+}_l\ket{+}_r+\alpha_+\beta_-\ket{+}_l\ket{-}_r+\alpha_-\beta_+\ket{-}_l\ket{+}_r-\alpha_-\beta_-\ket{-}_l\ket{-}_r\Big)+\nonumber\\
\fl\ket{2,1}_l\ket{3,1}_r\Big(\alpha_+\beta_+\ket{+}_l\ket{+}_r-\alpha_+\beta_-\ket{+}_l\ket{-}_r+\alpha_-\beta_+\ket{-}_l\ket{+}_r+\alpha_-\beta_-\ket{-}_l\ket{-}_r\Big)+\nonumber\\
\fl\ket{2,1}_l\ket{4,1}_r\Big(\alpha_+\beta_-\ket{+}_l\ket{+}_r-\alpha_+\beta_+\ket{+}_l\ket{-}_r+\alpha_-\beta_-\ket{-}_l\ket{+}_r+\alpha_-\beta_+\ket{-}_l\ket{-}_r\Big)+\nonumber\\
\fl\ket{3,1}_l\ket{1,1}_r\Big(\alpha_+\beta_-\ket{+}_l\ket{+}_r+\alpha_+\beta_+\ket{+}_l\ket{-}_r-\alpha_-\beta_-\ket{-}_l\ket{+}_r+\alpha_-\beta_+\ket{-}_l\ket{-}_r\Big)+\nonumber\\
\fl\ket{3,1}_l\ket{2,1}_r\Big(\alpha_+\beta_+\ket{+}_l\ket{+}_r+\alpha_+\beta_-\ket{+}_l\ket{-}_r-\alpha_-\beta_+\ket{-}_l\ket{+}_r+\alpha_-\beta_-\ket{-}_l\ket{-}_r\Big)+\nonumber\\
\fl\ket{3,1}_l\ket{3,1}_r\Big(\alpha_+\beta_+\ket{+}_l\ket{+}_r-\alpha_+\beta_-\ket{+}_l\ket{-}_r-\alpha_-\beta_+\ket{-}_l\ket{+}_r-\alpha_-\beta_-\ket{-}_l\ket{-}_r\Big)+\nonumber\\
\fl\ket{3,1}_l\ket{4,1}_r\Big(\alpha_+\beta_-\ket{+}_l\ket{+}_r-\alpha_+\beta_+\ket{+}_l\ket{-}_r-\alpha_-\beta_-\ket{-}_l\ket{+}_r-\alpha_-\beta_+\ket{-}_l\ket{-}_r\Big)+\nonumber\\
\fl\ket{4,1}_l\ket{1,1}_r\Big(\alpha_-\beta_-\ket{+}_l\ket{+}_r+\alpha_-\beta_+\ket{+}_l\ket{-}_r-\alpha_+\beta_-\ket{-}_l\ket{+}_r+\alpha_+\beta_+\ket{-}_l\ket{-}_r\Big)+\nonumber\\
\fl\ket{4,1}_l\ket{2,1}_r\Big(\alpha_-\beta_+\ket{+}_l\ket{+}_r+\alpha_-\beta_-\ket{+}_l\ket{-}_r-\alpha_+\beta_+\ket{-}_l\ket{+}_r+\alpha_+\beta_-\ket{-}_l\ket{-}_r\Big)+\nonumber\\
\fl\ket{4,1}_l\ket{3,1}_r\Big(\alpha_-\beta_+\ket{+}_l\ket{+}_r-\alpha_-\beta_-\ket{+}_l\ket{-}_r-\alpha_+\beta_+\ket{-}_l\ket{+}_r-\alpha_+\beta_-\ket{-}_l\ket{-}_r\Big)+\nonumber\\
\fl\ket{4,1}_l\ket{4,1}_r\Big(\alpha_-\beta_-\ket{+}_l\ket{+}_r-\alpha_-\beta_+\ket{+}_l\ket{-}_r-\alpha_+\beta_-\ket{-}_l\ket{+}_r-\alpha_+\beta_+\ket{-}_l\ket{-}_r\Big)\bigg).\nonumber\\
\end{eqnarray}
We now perform a measurement of the state of the two molecules in the $\ket{J,|m|}$ basis. The state of the quasi hidden degree of freedom of the two molecules is thereby projected onto the expression in the parentheses on one of the 16 lines of Eq.~\ref{finalstate}. As can be seen, each possible final state consists of a controlled $z$ gate performed on the the state $\ket{\alpha}_l\ket{\beta}_r$ followed by $\pi$ rotations on the two qubits, precisely as predicted.

The final step is to perform the required single qubit rotations on the quasi-hidden qubits. Specifically, we require the ability to add a $\pi$ phase shift to the state $\ket{+}$ relative to the state $\ket{-}$ and the ability to swap the states $\ket{+}$ and $\ket{-}$. We assume the rest of the molecule to be in a state $\ket{J,1}$. The first operation can be achieved, for example, by driving a $2\pi$ rotation between the states $\ket{J,1,+}$ and $\ket{J+1,2,+}$. The second operation can be achieved, for example, by driving a $2\pi$ rotation between the states $\ket{J,1}$ and $\ket{J,0}$. The state $(\ket{J,1,+}-\ket{J,1,-})/\sqrt{2}$ thereby remains unaffected, where the state $(\ket{J,1,+}+\ket{J,1,-})/\sqrt{2}$ acquires a $\pi$ phase shift. The states $\ket{J,1,+}$ and $\ket{J,1,-}$ are thereby interchanged.

\section{Conclusion}

In summary, we have explored the possibility of using molecules with a quasi-hidden degree of freedom for quantum information processing. Some of the benefits of using a quasi-hidden degree of freedom have already been demonstrated experimentally in a parallel paper~\cite{Loew23}. This includes the ability to observe long ($\sim100\mu s$) coherences in the hidden degree of freedom in a highly noisy environment. This could easily be extended by many orders of magnitude.

Our paper provides multiple opportunities for future work. First, various details need to be worked out to apply the ideas to quasi-hidden degrees of freedom other than opposite rotation state pairs. Second, while we have explored in detail how the quasi-hidden degree of freedom can be used once entanglement between the non-hidden degrees of freedom and an external system is achieved, we have only sketched how to achieve the entanglement with the external system in the first place. Here, detailed concepts could be developed for achieving high fidelity entanglement as discussed based on heralded entanglement or entanglement distillation. 
Finally, tremendous opportunity exists to explore the presented ideas experimentally.

\ack
Thanks to Florian Jung, Maximilian L\"ow, and Andreas Reiserer for helpful discussions.
\vspace{3mm}

\bibliographystyle{unsrt}

\end{document}